\documentclass[a4paper]{article}
\usepackage{graphicx}
\usepackage{amsmath}

\begin{document}

\begin{center}
{\bf \Large
The Heider balance - a continuous approach
}\\[5mm]

{\large
Krzysztof Ku{\l}akowski, Przemys{\l}aw Gawro{\'n}ski and Piotr Gronek
}\\[3mm]

{\em
Faculty of Physics and Applied Computer Science,
AGH University of Science and Technology\\
al. Mickiewicza 30, PL-30059 Krak\'ow, Poland
}

\bigskip
{\tt kulakowski@novell.ftj.agh.edu.pl}

\bigskip
\today
\end{center}

\begin{abstract}
The Heider balance (HB) is investigated in a fully connected graph of $N$ nodes. The links 
are described by a real symmetric array $r(i,j)$, $i,j=1,...,N$. In a social group, 
nodes represent group members and links represent relations between them, positive (friendly)
or negative (hostile). At the balanced state, $r(i,j)r(j,k)r(k,i)>0$ for all the triads $(i,j,k)$. 
As follows from the structure theorem of Cartwright and Harary, at this state the group
is divided into two subgroups, with friendly internal relations and hostile relations 
between the subgroups. Here the system dynamics is proposed to be
determined by a set of differential equations, $\bf \dot{r}=r\cdot r$. The form of equations 
guarantees that once HB is reached, it persists. Also, for $N=3$ the dynamics 
reproduces properly the tendency of the system to the balanced state. The equations are solved
numerically. Initially, $r(i,j)$ are random numbers distributed around zero with a symmetric 
uniform distribution of unit width.  Calculations up to $N=500$ show that HB is always 
reached. Time $\tau(N)$ to get the balanced state varies with the system size $N$ as $N^{-1/2}$. 
The spectrum of relations, initially narrow, gets very wide near HB. This means that the 
relations are strongly polarized. In our calculations, the relations are limited to a given range 
around zero. With this limitation, our results can be helpful in an interpretation of some
statistical data.
 
\end{abstract}

\noindent

\noindent
{\em Keywords:}  sociophysics; interpersonal relations; fully connected network

\section{Introduction}

The purpose of this work is to develop a mathematical formulation of the time evolution
of social relations. Theory of this problem was initialized in 40's by Fritz Heider 
\cite{hei1,hei2,kad}. As we will see below, some ingredients of the 
Heider approach are present in the Sznajd model of convincing people, designed in 2000 
\cite{szn, sta}. 

The Heider balance (HB) is a concept in social psychology \cite{ntc}. For the purposes of 
this text, HB can be summarized as follows. A set of $N$ nodes, which represent a group 
of its human members, form a fully connected graph, i.e. each pair of nodes is linked. 
Each link is represented by an element $r_{i,j}$ of a symmetric matrix; it takes a value 
$\pm 1$. (Our discussion here does not include the possibility when $r_{i,j}=0$, what is
 more complex.) This element is a measure of the relation between $i$ and $j$; it is 
positive if they are friends,  negative in the opposite. In the HB state, the product 
$r_{i,j}r_{j,k}r_{k,i}$ is positive for each triad $(i,j,k)$ of nodes. This means 
that either all $r's$ in the triad are positive or exactly two of them are negative. 

This kind of balance or equilibrium of a social system has been designed to reproduce the 
human tendency to preserve a cognitive consistency of hostility and friendship. The principle 
is simple: 'my friend's friend is my friend, my friend's enemy is my enemy, my enemy's 
friend is my enemy, my enemy's enemy is my friend' \cite{hei2}. Once the balance is not 
present, there is a kind of tension in the group members' minds which can eventually 
lead to changes in their opinions. Once the balance is present, it appears to be stable,
 because there is no cognitive dissonance which could change the state. More information 
and references can be found in \cite{dor2}.

It appears that HB is equivalent to a partition of the graph into two separate subgraphs. 
Within each of them, all links are positive (i.e. $r_{i,j}>0$ if both nodes $i$ and $j$ 
belong to the same subgraphs). On the contrary, links between nodes belonging to different
 subgraphs are negative. This statement is the content of the so-called structure theorem 
(\cite{har,dor1}). In social language, HB means that there are two antagonistic 
groups, with perfect accordance within them and pure hostility between them. The effect 
has been termed 'social mitosis' \cite{wang}. As a special case, HB includes also states 
where all links are positive and there is only one subgraph, identical with the whole graph. 
In recent computational experiment, several graphs were analyzed with different initial 
distribution of the signs of links. There, HB was identified as a final state in all 
investigated cases \cite{wang}. The dynamics included changes of sign of links contained 
by unbalanced triads. As far as we know, there is no proof that this dynamics always leads 
to HB \cite{wang}. It is clear that HB is stable with respect to this kind of evolution: 
once HB is reached, the system remains unchanged.

Our aim here is to use real numbers instead of $\pm 1$ to describe the opinion distribution 
and its dynamics. There are heavy arguments to introduce this modification. First, as noted 
in Ref. \cite{wang}, the relations between human beings vary in strength and not only in 
sign. Moreover, techniques of measurements of this strength are well established in sociology 
\cite{ntc}. In particular, the famous Bogardus scale of social distance \cite{bog1}
 seems to be appropriate for our purposes. (For its recent application see e.g. Ref.\cite{kleg}.) 
 The only modification needed is to convert it by decreasing function $f(x)$ as to get a
 positive number (say, $R$) for shortest social distance $(f(1)=R)$, and a negative number 
(say, $-R$) for the largest social distance $s_{max}$ ($f(s_{max})=-R$ or so). Second, dynamics 
expressed
 in terms of differential equations allows to capture different timescales of the processes,
 which are present in social groups. In Ref. \cite{dor2}, three of them are named: reciprocity,
 transitivity and balance. Although here we deal with the balance only (which is supposed to
 be the slowest), the modification is promising for future generalizations of the model. Still, 
the velocity of time evolution of a particular link can depend on the state of the 
network and in general it varies in time. Third, the question 'which link changes as first?' 
ubiquitous in discrete simulations, is evaded; in the continuous formulation all links 
are allowed to evolve simultaneously. 

\section{Equations}

In the original discrete picture, opinion of $i$ on $j$ is influenced by $k$ as follows. An 
example of imbalanced triad is, when there are two positive links and one negative. This can 
mean that, example giving, $i$ likes both $j$ and $k$ (two positive links) but $k$ dislikes
 $j$. For $i$, there is a cognitive dissonance. It can be removed by rejecting either $i$ 
or $j$. In both cases, the division for good and bad people ($(ij)$ vs $(k)$ or $(ik)$ vs 
$(j)$) becomes consistent. 

Instead, we propose a set of differential equations

\begin{equation}
\frac{dr(i,j)}{dt} = \sum_k r(i,k) r(k,j)
\end{equation}

For a triad $(x,y,z)$ of three nodes this set is equivalent to three nonlinear equations

\begin{eqnarray} 
\frac{dx}{dt}& = yz \\
\frac{dy}{dt}& = zx \\
\frac{dz}{dt}& = xy
\end{eqnarray}

The set of fixed points consists three coordinate axes: ($x=y=0$) plus ($y=z=0$) plus ($z=x=0$). 
We ask if these fixed points are stable. Consider, for example, the OX axis where ($y=z=0$). 
The Jacobian at the fixed point is

\begin{displaymath}
\mathbf{J}=
\left( \begin{array}{ccc}
0&0&0 \\
0&0&x \\
0&x&0
\end{array}\right)
\end{displaymath}
with the eigenvalues (0,+x,-x). We know that the stability condition demands all eigenvalues 
to be negative \cite{glen}. It is clear that the fixed points are not stable. The condition
 of HB reduces to $xyz>0$, what is fulfilled in four out of eight parts of space of coordinates
 $(x,y,z)$. This condition, once true, remains true forever: either there are no negative bonds 
and all time derivatives are positive, or two bonds are negative and their time derivatives 
are negative as well. 

For larger $N$ the method  of inspection ceases to be simple, and therefore we rely on numerical 
simulations. However, we can add a simple observation which is valid for any value of $N$.
Multiplying Eq. (1) by $r(i,j)$, we get on r.h.s. a sum on expressions $r(i,k)r(k,j)r(j,i)$,
each of them is positive at HB. On l.h.s, we get $\dot r^2(i,j)/2$. This means that 
$r(i,j)\dot r(i,j)>0$ at HB. Then at HB either $r(i,j)>0$ and $\dot r(i,j)>0$, or $r(i,j)<0$ and
$\dot r(i,j)<0$: positive links increase, negative links decrease. This property, once true,
cannot change; therefore HB is stable. We note that both sides of Eq.(1) can be positive 
before HB is reached. We have seen a numerical example of such a state. 

Basic result of this work is to demonstrate numerically that the dynamics given by Eq.1 leads 
to HB in all investigated cases. This result and the method is analogous to those of Ref. 
\cite{wang}: we start from a random initial state and we average the obtained results over
 a reasonable number of graphs of the same size $N$. What is calculated is the time to get
 HB, as dependent on the number $N$ of nodes. We apply the Runge-Kutta method of 4-th order,
with adjusting the length of timesteps. 

We are also interested on the time evolution of the probability distribution of links $r_{i,j}$.
 However, near HB the matrix
 elements increase without limits. We believe that this is not so in the psychosociological 
reality, where extreme opinions do not spread just because we prefer to be considered
 as civilized people. In other words, the Bogardus scale remains finite at both ends. 
It is reasonable to introduce this limitation into the equations. We use an envelope

\begin{equation}
G(r;R)=1-\frac{r^2(i,j)}{R^2}
\end{equation}
as a multiplicative  factor in the equations of motion, which take the form

\begin{equation}
\frac{dr(i,j)}{dt} = G(r;R) \sum_k r(i,k) r(k,j)
\end{equation}

We guess that the dynamics of getting HB is not influenced if $R$ is large enough.

\section{Results}

In Fig. 1 we show the time $T(N)$ of getting HB as dependent on the system size $N$. 
The log-log plot reveals that above $N=100$, the slope of the curve is consistent with 
the law $T\propto N^{-1/2}$.  These results are obtained both for $R=5.0$ and infinity; 
in the latter case, $G(r;R)=1$. We deduce that the limitation of $R=5.0$ does not 
influence the dynamics until HB is reached.

\begin{figure}
\begin{center}
\includegraphics[angle=-90,width=.8\textwidth]{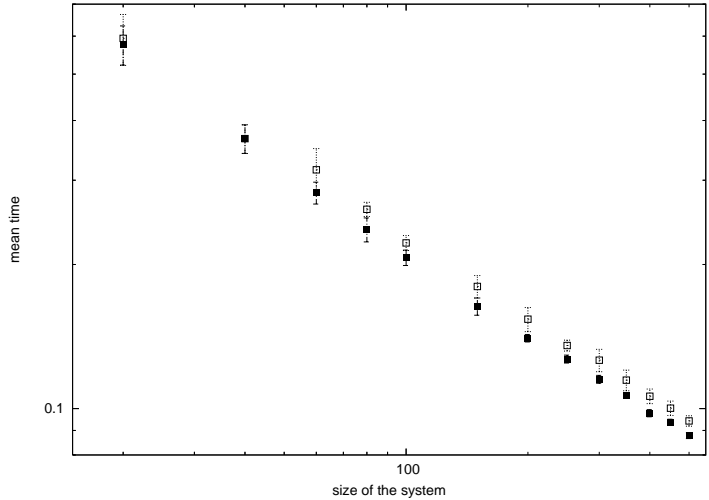}
\caption{Time to reach HB as dependent on the system size $N$ for $R=5.0$ (open squares)
and $R=\infty$ (black squares).}
\label{t-N}
\end{center}
\end{figure}

\begin{figure}
\begin{center}
\includegraphics[angle=-90,width=.8\textwidth]{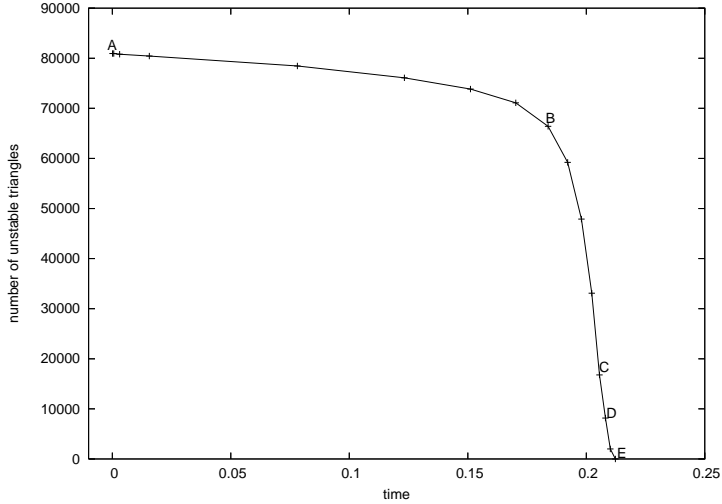}
\caption{Departure from HB as a number of triangles, against time, for one initial 
distribution of $r(i,j)$. For points marked with labels A-E, the distributions of $r(i,j)$
for $R=5.0$ are shown in Fig. 3.}
\label{n-t}
\end{center}
\end{figure}

\begin{figure}
\begin{center}
a)\includegraphics[angle=-90,width=.45\textwidth]{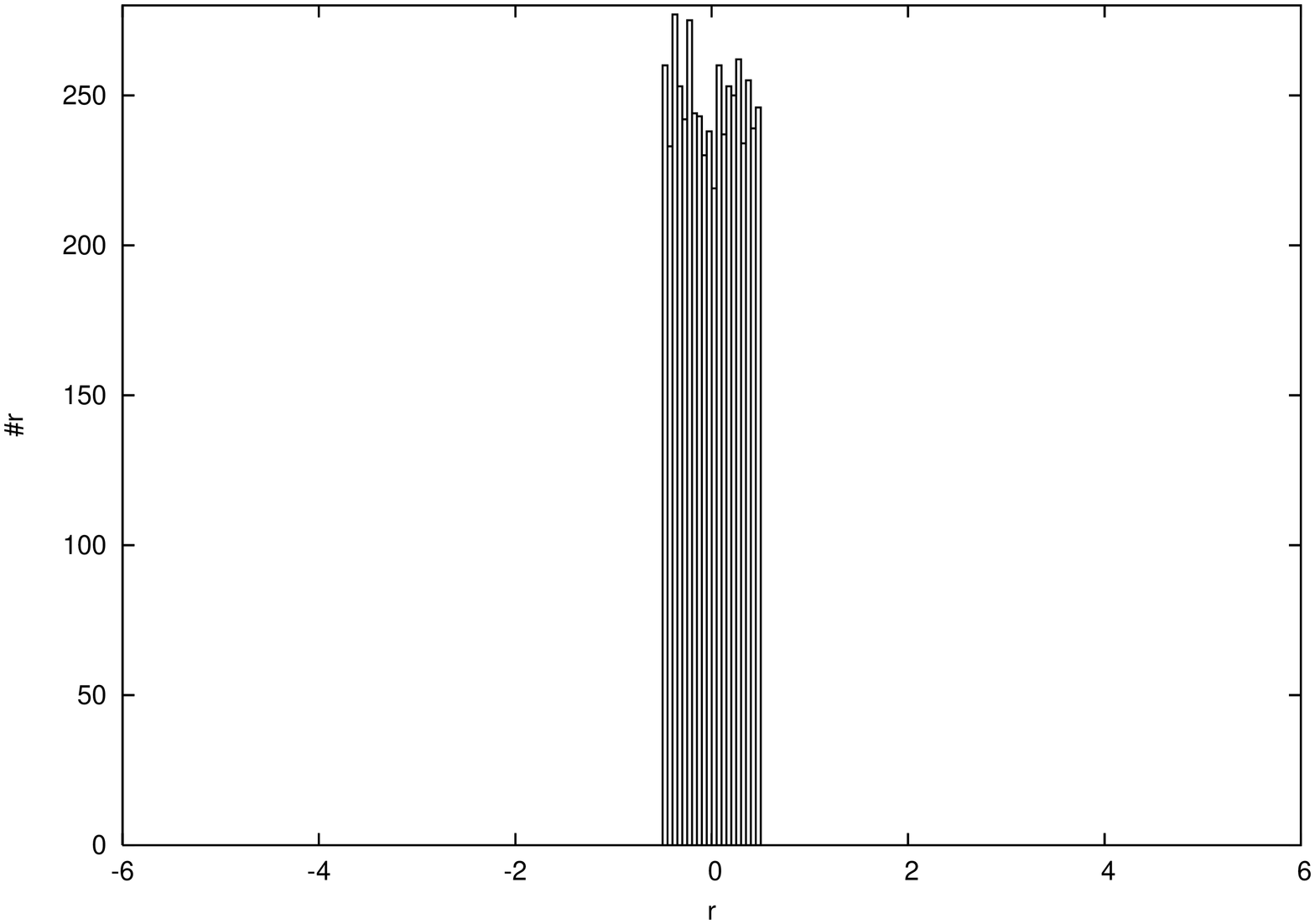}
b)\includegraphics[angle=-90,width=.45\textwidth]{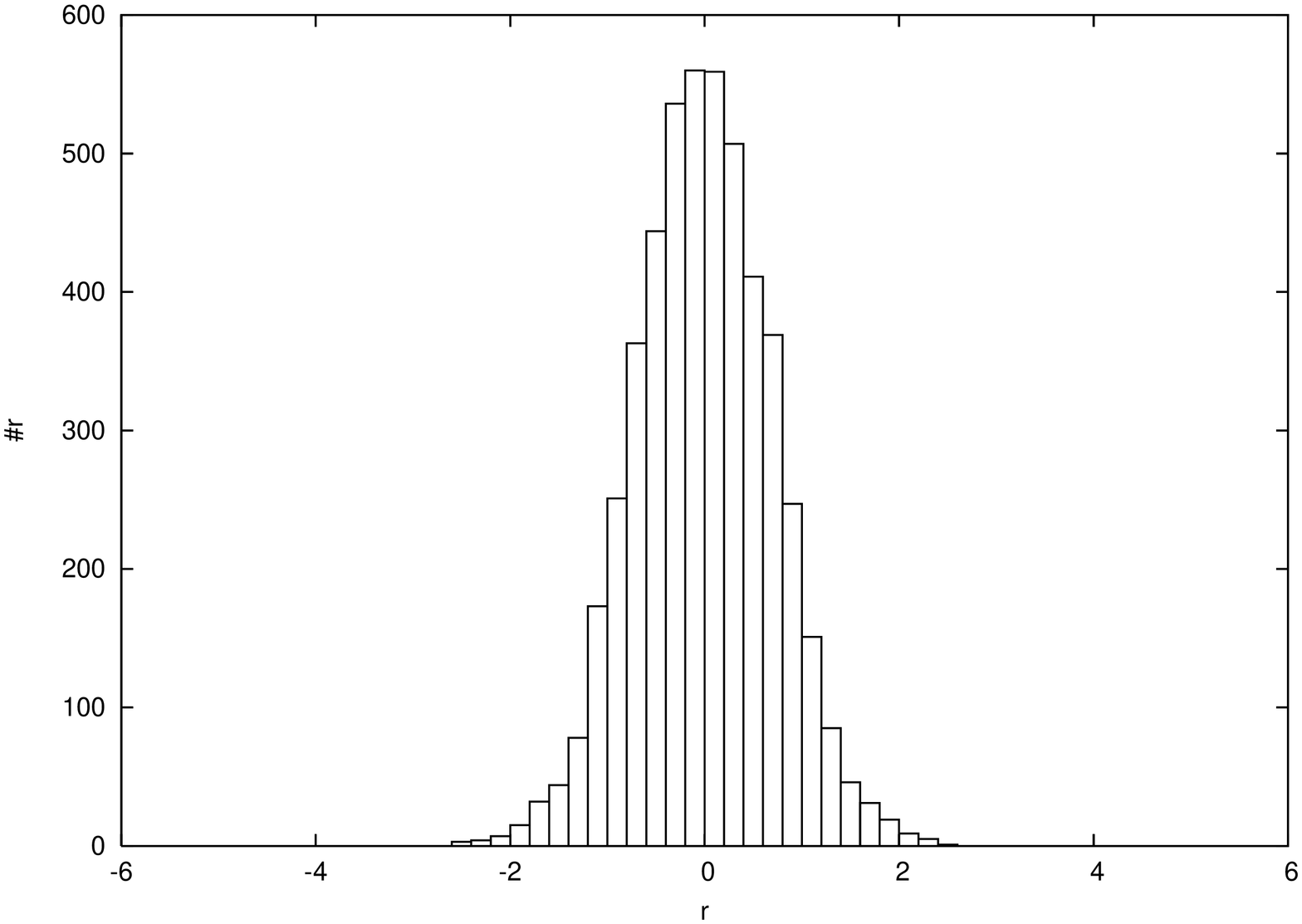}
c)\includegraphics[angle=-90,width=.45\textwidth]{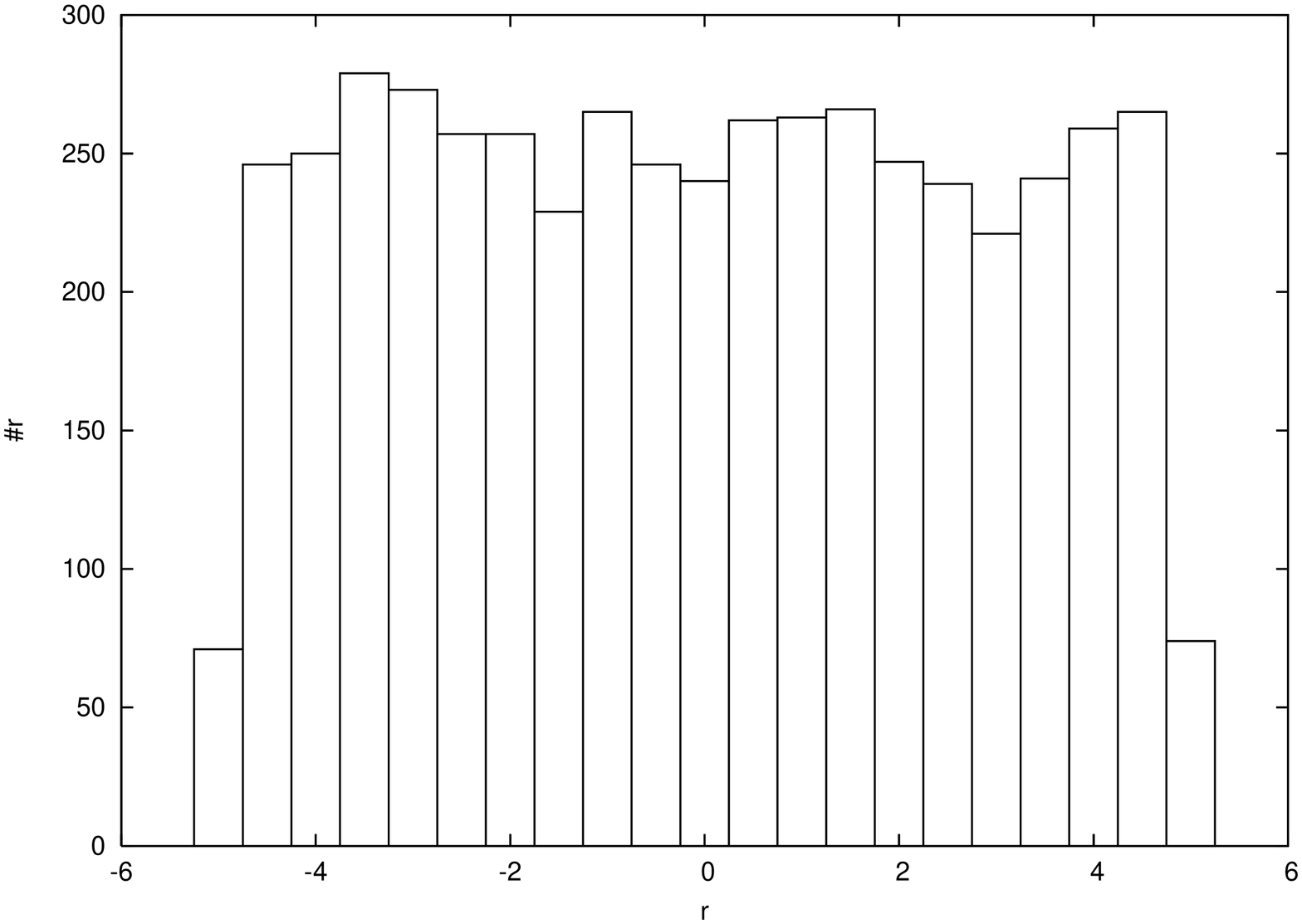}
d)\includegraphics[angle=-90,width=.45\textwidth]{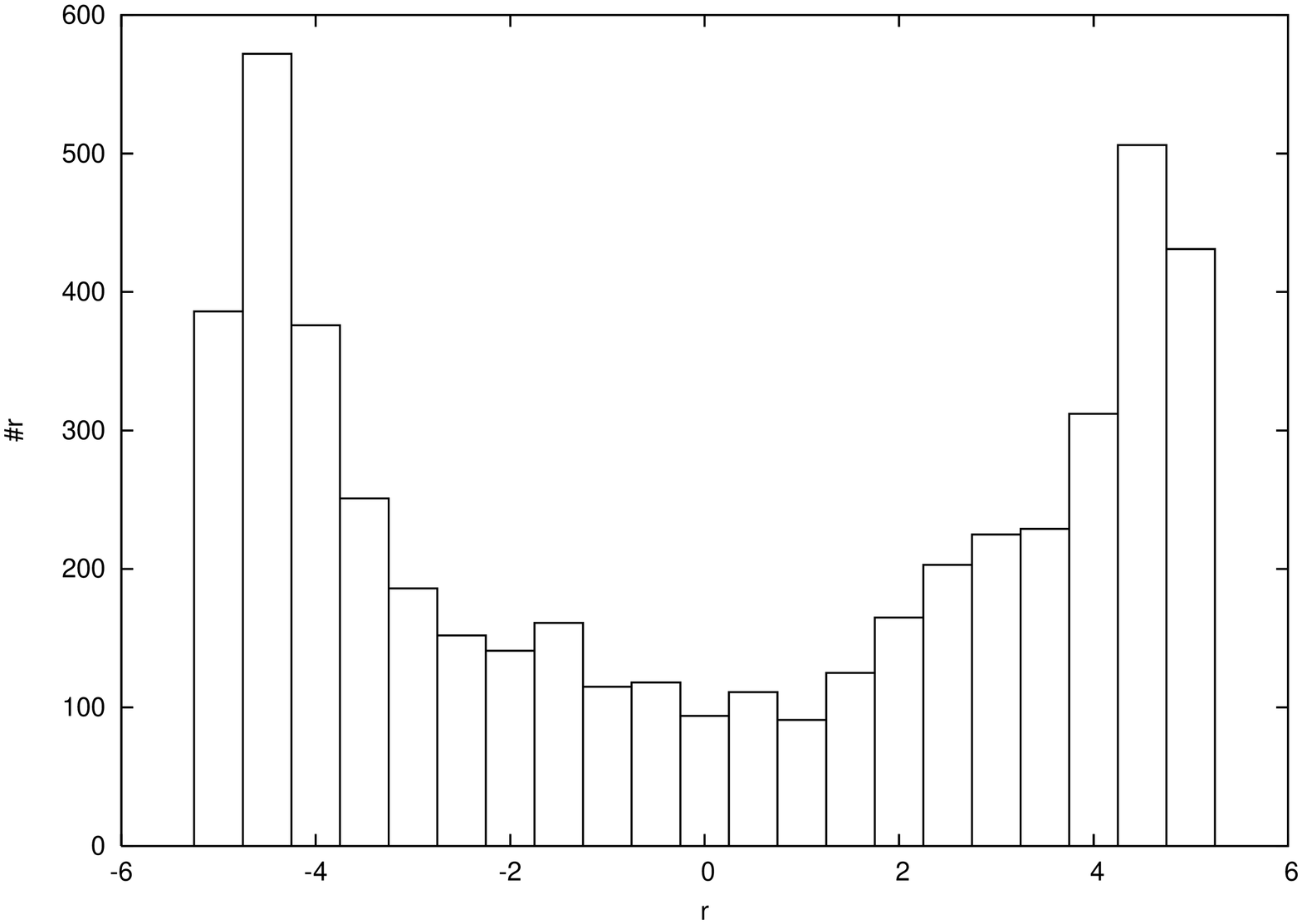}
e)\includegraphics[angle=-90,width=.45\textwidth]{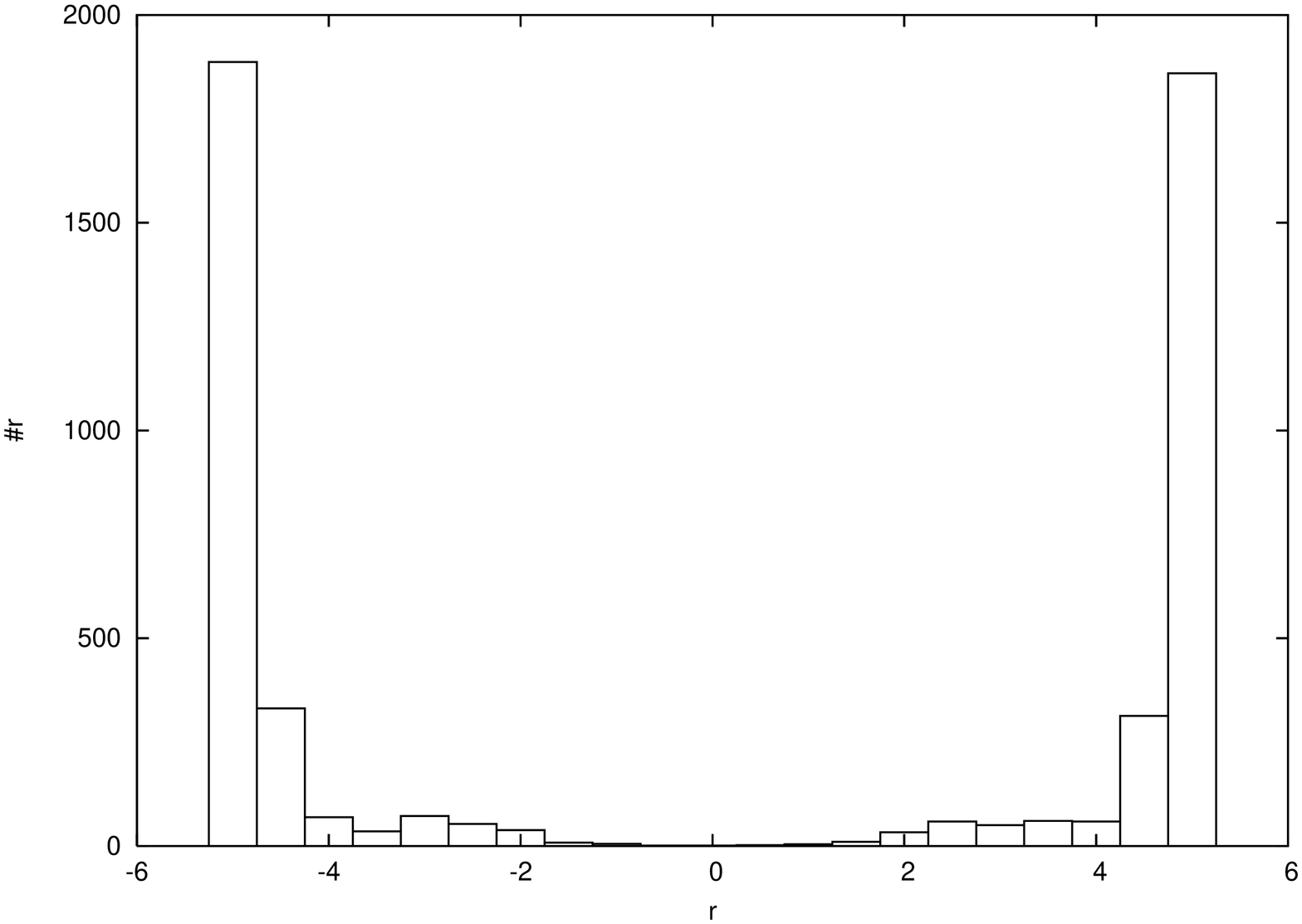}
\caption{Distribution of $r(i,j)$ at point A, B, C, D, E.}
\end{center}
\end{figure}

\begin{figure}
\begin{center}
\includegraphics[angle=-90,width=.8\textwidth]{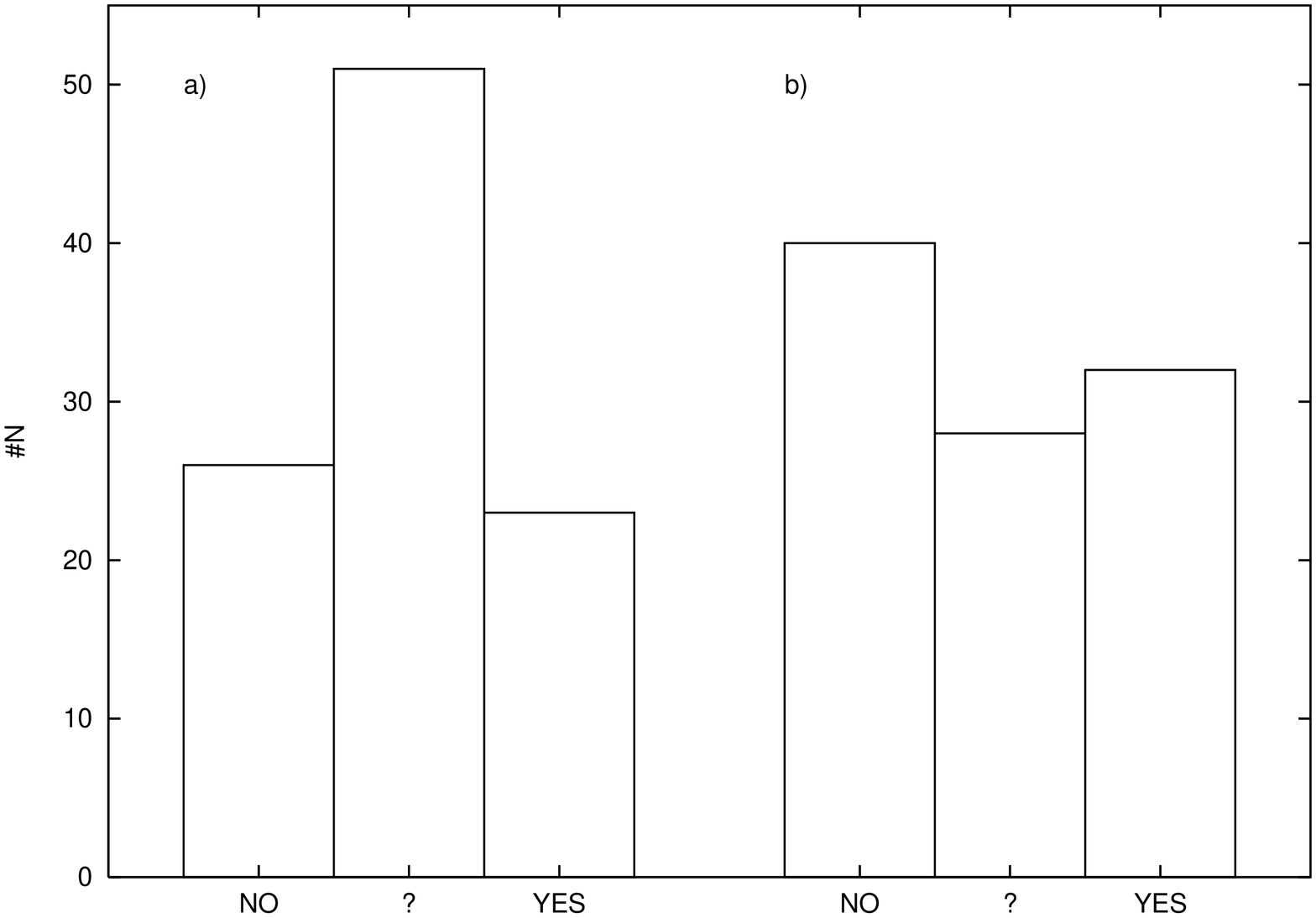}
\caption{Distribution of opinions of the anti-communist law (a) in May 99, (b) in November 99.
The data are averaged over four political parties, important in 1999 (AWS, UW, PSL, SLD) and 
voices of indifferent people, with equal weights \cite{cbos1}. }
\label{p(r)-exp}
\end{center}
\end{figure}

On the contrary, it is more convenient to evaluate the statistics of $r(i,j)$ when its 
size is limited. Also, as remarked in the preceding section, a limitation of $r(i,j)$ is 
justified from the basic point of view. Then we calculate the time dependence of the 
distribution of $r(i,j)$ for the case when $R=5.0$. The results are shown in Fig.2.
The distribution of the matrix elements $r(i,j)$ at different stages of the process
is shown in Figs.3-7.

The above value $-1/2$ of the exponent is easy to be explained as a superposition of two
 factors. As it follows from Eq.1, the velocity of the variation of the matrix elements 
$r(i,j)$ is a summation over 'third' nodes of the triads. As such, it increases linearly 
with $N$. On the other hand, r.h.s of Eq. (1) is composed from random numbers with initial average 
equal zero. The deviation from the average is known to decrease with $N$ as $N^{-1/2}$.
 The velocity as a product increases as $N^{1/2}$, and the time $T(N)$ is its inverse. 
This argument on the time $T$ as an inverse of the velocity holds when the distribution
 of the matrix elements is symmetric around zero. For asymmetric distributions the obtained 
time $T$ is shorter. The system goes to HB even if all initial $r$'s are negative. In this 
case the distribution is continuously shifted in time towards positive values.

\section{Discussion}

In fully connected networks, at most two subgraphs can appear at the Heider balance. 
We note that this is a consequence of the assumption that a triad with all links negative 
remains unbalanced \cite{har,dor1}.
However, it is easy to imagine that instead of a fully connected graph we have 
a shell structure where there is a number of 
subgraphs connected in a chain one to another with negative links. In this 
case, each subgraph can reach HB separately. In social reality, we can rarely state that a 
given social system is balanced; the polarization of opinions seems to be much better
to investigate and predict conflicts. The most important goal of this
work is a conclusion, is that the removing of a cognitive dissonance is accompanied 
with the polarization of opinions. This result cannot be obtained if opinions are 
represented merely by $\pm 1$. 

This polarization is due both to the positive interaction within the subgraph ('my friend's 
friend is my friend') and the negative interaction between the subgraphs (' my enemy's enemy 
is my friend'). Once HB is reached, each opinion - negative and positive - is enhanced by 
the interaction with other people. In the sociological literature, the effect is known as 
'echo hypothesis' \cite{burt}. Eq. (1) can be treated as its mathematical formulation.
This positive feedback between neighboring links is to some extent reflected also in the Sznajd 
model \cite{szn}, where a consent of opinions of two people influences their neighbours.
 
In our approach, the dynamics is local: evolution of each bond $r(i,j)$ is determined 
solely by the state of triads which contain the nodes $i$ and $j$. This is close to 
existing social systems, where people do not count the balanced triads in the whole network. 
On the other 
hand, in reality people change their opinions on other people classifying them in 
categories, along social, political or ethnic criteria. These cathegories are known 
to play major role in conflict emerging. This is a difference between this approach and 
other models of opinion dynamics, as the 
Sznajd model \cite{szn}, where opinions on some general ideas are considered. However, 
these general ideas could deal as well with the above mentioned classifying criteria. For example,
the idea of racism influences relations between people of different races, etc.

The goal of this work is to improve the agreement between the Heider model of removing
the cognitive dissonance and the mathematical realization of this model. Such aim
is less ambitious than a proof that the model is true by comparing the its results 
with reality, what is always a formidable task in social sciences \cite{col}. However, 
for an illustration of the idea
we have found an example in recent history of Poland, where the public opinion was strongly
coupled to relations between actors at the political scene. For some time, curricula of public
people are investigated by law to check if they collaborated with the communist secret service
before 1989. Analysis of opinions on this law displayed no great interest in May 1999 \cite{cbos1}.
However, in June 1999 the law was applied to our former Prime Minister. Then people  
started to wonder what their political friends know about them, who is reliable etc. 
Although the discussion on the law remained general and abstract, interpersonal relations were 
influenced by fear. The resulting polarization of opinions is shown in Fig. 4. We note 
that in other cases, as the war in Iraq \cite{cbos2} or the Polish membership in EU \cite{cbos3}, 
the splitting is not observed; instead, the maxima of the appropriate distributions shift 
in time from 'YES' to 'NO'. In these cases, a given opinion does not imply a personal solidarity 
with any group.

\bigskip
{\bf Acknowledgements}. The authors are grateful to Dietrich Stauffer who planted 
sociophysics in our team.

\bigskip

\end{document}